%% file: alicepreprint_CDS.tex
\documentclass[ALICE,manyauthors]{cernphprep}

\usepackage[comma,square,numbers,sort&compress]{natbib}
\usepackage{hyperref}
\usepackage{lineno}
\usepackage{color}
\begin{document}%
%%%%%%%%%%%%%%%  Title page %%%%%%%%%%%%%%%%%%%%%%%%
\begin{titlepage}
\PHyear{2018}
\PHnumber{119}      % required, will be obtained from PH
\PHdate{9 May}  % required, will be obtained from PH
%

%%% Put your own title + short title here:
\title{Azimuthal anisotropy of heavy-flavour decay electrons in p--Pb collisions at $ \sqrt{s_{\rm{NN}}}=5.02$ TeV}
\ShortTitle{Azimuthal anisotropy of heavy-flavour decay electrons in p--Pb collisions}   % appears on right page headers

%%% Do not change the next lines
\Collaboration{ALICE Collaboration\thanks{See Appendix~\ref{app:collab} for the list of collaboration members}}
\ShortAuthor{ALICE Collaboration} % appears on left page headers, do not change

\begin{abstract}
Angular correlations between heavy-flavour decay electrons and charged particles at mid-rapidity ($|\eta| < 0.8$) are measured in p--Pb collisions at $\sqrt{s_{\rm{NN}}} = 5.02$ TeV. The analysis is carried out for the 0--20\% (high) and 60--100\% (low) multiplicity ranges. The jet contribution in the correlation distribution from high-multiplicity events is removed by subtracting the distribution from low-multiplicity events. An azimuthal  modulation remains after removing the jet contribution, similar to previous observations in two-particle angular correlation measurements for light-flavour hadrons. A Fourier decomposition of the modulation results in a positive second-order coefficient ($v_{2}$) for heavy-flavour decay electrons in the transverse momentum interval $1.5 < p_{\rm{T}} < 4$~GeV/$c$ in high-multiplicity events, with a significance larger than $5\sigma$.  The results are compared with those of charged particles at mid-rapidity and of inclusive muons at forward rapidity. The $v_2$ measurement of open heavy-flavour particles at mid-rapidity in small collision systems could provide crucial information to help interpret the anisotropies observed in such systems.
%This is the first $v_2$ measurement of open heavy-flavour particles at mid-rapidity in small collision systems, which could provide crucial information to help interpret the anisotropies observed in such systems.
\end{abstract}
\end{titlepage}
\setcounter{page}{2}

\input{Introduction}

\input{Analysis}

\input{Conclusion}

%%%%% acknowledgements
\newenvironment{acknowledgement}{\relax}{\relax}
\begin{acknowledgement}
\section*{Acknowledgements}
\input{fa_2018-05-09.tex}    %%%%%%% done by webmaster team
\end{acknowledgement}

%%%%%%%% Bibliography (In case of using bibtex generate the bbl requested by arXiv)
%\clearpage
\bibliographystyle{utphys}   % Remember we use title in the biblio
\bibliography{bibliography.bib}

%%%%%%%%% appendix with author list
\newpage
\appendix
\section{The ALICE Collaboration}
\label{app:collab}
\input{Alice_Authorlist_2018-May-01.tex}  %%%%%%% done by webmaster team
\end{document}

%% file: Introduction.tex
Two-particle angular correlations are a powerful tool to study the dynamical evolution of the system created in ultra-relativistic collisions of protons or nuclei. The differences in the azimuthal angle ($\Delta\varphi$) and in pseudorapidity ($\Delta\eta$) between a reference (``trigger'') particle and other particles produced in the event are considered. The typical shape of the correlation distribution features a near-side peak at $(\Delta\varphi, \Delta\eta) \sim (0,0)$, 
%from cases when the 
originating from cases in which the trigger particle is produced in a jet, and an away-side structure centered at $\Delta\varphi \sim \pi$ and extending over a wide pseudorapidity range, due to the recoil jet~\cite{Wang:1992db}.
In nucleus--nucleus collisions the correlation distribution also exhibits pronounced structures on the near- and away-side extending over a large $\Delta\eta$ region, commonly referred to as ``ridges''~\cite{Abelev:2009af}. 
They can be quantified by the $V_{\rm{n}\Delta}$ coefficient of a Fourier decomposition of the $\Delta\varphi$ distribution, which is performed after removing the jet contribution.
%They can be quantified via the coefficients $V_{\rm{n}\Delta}$ of the Fourier decomposition of the correlation distribution $\Delta\varphi$ projection, obtained after removing the jet contribution.
These coefficients can be factorised into single-particle coefficients $v_{\rm{n}}$ related to the azimuthal distribution of the particles with respect to the \textit{n-th} order symmetry planes~\cite{Aamodt:2011by}.
In non-central nucleus--nucleus collisions, the dominant coefficient is that of the second-order harmonic, referred to as elliptic flow ($v_{\rm{2}}$),
and its value is used to characterise the collective motion of the system.
The measurements are well described by models invoking a hydrodynamic expansion of the hot and dense medium produced in the collision. This translates the initial-state spatial anisotropy, due to the asymmetry of the nuclear overlap region, into a momentum anisotropy of the particles emerging from the medium~\cite{Qin:2010pf}. This collective motion is one of the important features of the Quark-Gluon Plasma (QGP) produced in such collisions. 

Surprisingly, the presence of similar long-range ridge structures and a positive $v_{\rm{2}}$ coefficient were also
observed for light-flavour hadrons in high-multiplicity proton--lead (p--Pb) collisions by the ALICE~\cite{Abelev:2012ola}, ATLAS~\cite{Aaboud:2016yar} and CMS~\cite{Chatrchyan:2013nka} collaborations at the LHC. The pattern of the $v_{\textrm{2}}$ coefficient as a function of the particle mass and transverse momentum is similar in p--Pb and Pb--Pb collisions~\cite{ABELEV:2013wsa, Khachatryan:2014jra}. The PHENIX and STAR collaborations at RHIC also measured a positive $v_{\rm{2}}$ coefficient for charged hadrons in high-multiplicity d--Au and $^3$He--Au collisions~\cite{Adare:2013piz,Adamczyk:2015xjc,Adare:2015ctn}. A near-side structure extended over a large $\Delta\eta$ range was also reported for high-multiplicity proton--proton (pp) collisions by the CMS~\cite{Khachatryan:2010gv} and ATLAS~\cite{Aaboud:2016yar} collaborations. The interpretation of a positive $v_{\rm{2}}$ in these small collision systems is currently highly debated~\cite{Loizides:2016tew}. 
One possible interpretation is based on collective effects induced by a hydrodynamical evolution of the particles produced in the collision~\cite{Werner:2010ss, Deng:2011at}. Other approaches include mechanisms involving initial-state effects, such as gluon saturation within the Color-Glass Condensate effective field theory~\cite{Dusling:2013qoz, Bzdak:2013zma}, or final-state colour-charge exchanges~\cite{Dumitru:2013tja,Wong:2011qr}.

Because of their large masses, heavy quarks are produced in hard scattering processes during the early stages of hadronic collisions~\cite{Andronic:2015wma}. In heavy-ion collisions, the elliptic flow of charm mesons~\cite{Adamczyk:2017xur,Acharya:2017qps, Abelev:2014ipa, Sirunyan:2017plt} and heavy-flavour decay leptons \cite{Adare:2010de,Adamczyk:2014yew,Adam:2015pga,Adam:2016ssk,Aaboud:2018bdg} was found to have similar magnitude as that of charged particles~\cite{Abelev:2012di}, dominated by light-flavour hadrons. A search for a non-zero $v_{\rm{2}}$ in the correlation pattern of heavy-flavour particles in high-multiplicity p--Pb collisions could provide further insight on the initial- and final-state origin of the anisotropies in this collision system, helping in constraining the models that describe the ridge structures. The production mechanisms of heavy quarks, involving a large squared four-momentum transfer, are also different from those of light-flavour quarks. This gives the possibility to investigate whether the onset of the anisotropy of the particle azimuthal distribution is affected by the details of hard scattering and fragmentation processes.

In this letter, we present the %first 
measurement of $v_{\rm{2}}$ for open heavy-flavour particles at mid-rapidity in high-multiplicity p--Pb collisions at $\sqrt{s_{\rm{NN}}} = 5.02$ TeV via azimuthal correlations of electrons from charm- and beauty-hadron decays, and charged particles.
This result complements our previous studies of hidden-charm particles based on the measurement of the correlations between J/$\psi$ mesons at forward rapidity and charged particles at mid-rapidity in high-multiplicity p--Pb collisions at $\sqrt{s_{\rm{NN}}} =$ 5.02 TeV and 8.16 TeV, which found evidences for a positive $v_{\rm{2}}$ of J/$\psi$ mesons~\cite{Acharya:2017tfn}. 
A positive $v_{\rm{2}}$ of muons at forward and backward rapidity, which are predominantly produced by heavy-flavour decays for transverse momentum $(p_{\rm T})$ greater than 2 GeV/$c$, was also measured in high-multiplicity p--Pb collisions at $\sqrt{s_{\rm{NN}}} =$ 5.02 TeV~\cite{Adam:2015bka}. Similar indications of positive $v_{\rm{2}}$ were also reported at mid-rapidity in high-multiplicity p--Pb collisions at $\sqrt{s_{\rm{NN}}} =$ 8.16 TeV for D$^{0}$ mesons by the CMS ~\cite{Sirunyan:2018toe} collaboration and in preliminary results for D$^{*+}$ mesons~\cite{ATLAS-CONF-2017-073} and heavy-flavour decay muons~\cite{ATLAS-CONF-2017-006} by the ATLAS collaboration. 

%Similar indications of positive $v_{\rm{2}}$ were also reported in preliminary results at mid-rapidity in high-multiplicity p--Pb collisions at $\sqrt{s_{\rm{NN}}} =$ 8.16 TeV for D$^{*+}$ mesons~\cite{ATLAS-CONF-2017-073} and heavy-flavour decay muons~\cite{ATLAS-CONF-2017-006} by the ATLAS collaboration and for D$^{0}$ mesons by the CMS~\cite{CMS-PAS-HIN-17-003} collaboration.

%The ALICE collaboration has also measured angular correlations of muons at forward rapidity with charged particles at mid-rapidity in high-multiplicity p--Pb collisions at $\sqrt{s_{\rm{NN}}} =$ 5.02 TeV~\cite{Adam:2015bka}. 
%A positive $v_{\rm{2}}$ was observed for muons, which are predominantly produced by heavy-flavour decays for transverse momentum $(p_{\rm T})$ greater than 2 GeV/$c$. 
%Preliminary measurements from the ATLAS collaboration show a positive $v_{2}$ for heavy-flavour decay muons at mid-rapidity in high-multiplicity p--Pb collisions at $\sqrt{s_{\rm{NN}}} =$ 8.16 TeV~\cite{ATLAS-CONF-2017-006}. Similar indications were also reported in preliminary results for D$^{0}$ and D$^{*+}$ mesons at mid-rapidity by the CMS~\cite{CMS-PAS-HIN-17-003} and ATLAS~\cite{ATLAS-CONF-2017-073} collaborations, respectively.

%% file: Analysis.tex
%Data taken information
The data sample used for the analysis was collected by the ALICE experiment~\cite{Aamodt:2008zz,Abelev:2014ffa} in 2016 during the LHC p--Pb run at $\sqrt{s_{\rm{NN}}} = 5.02$ TeV. The center-of-mass reference frame of the nucleon--nucleon collision was shifted in rapidity by 0.465 units in the proton-going direction with respect to the laboratory frame. The events were recorded using a minimum-bias trigger, which required coincident signals in the two scintillator arrays of the V0 detector, covering the full azimuthal angle in the pseudorapidity ($\eta$) ranges $2.8 < \eta < 5.1$~(V0-A) and $-3.7 < \eta < -1.7$~(V0-C). Together with the V0 information, signals from the two Zero-Degree Calorimeters were used %information are used offline 
to reject beam-induced background. 
Only events with a primary vertex reconstructed within $\pm$10 cm from the centre of the detector along the beam axis were accepted. About $6 \times 10^{8}$ events, corresponding to an integrated luminosity of $L_{\rm{int}} = 295~\pm~11~\mu\rm{b}^{-1}$, were obtained after these selections. Only events in high- (0--20\%) and low-multiplicity (60--100\%) classes, evaluated using the signal amplitude in the V0-A detector~\cite{Adam:2014qja}, were considered.

%Electron and charged hadrons track cuts
Electrons with transverse momentum ($p_{\rm T}^{\rm e}$) in the interval $1.5 < p_{\rm T}^{\rm e} < 6$ GeV/$c$ and $|\eta| < $ 0.8 (corresponding to $-1.26 < y_{\rm{cms}}^{\rm e} < 0.34$, where $y_{\rm{cms}}^{\rm e}$ is the electron rapidity in the center-of-mass reference frame) were selected using similar criteria as discussed in~\cite{Adam:2015qda}. 
Charged tracks were reconstructed using the Inner Tracking System (ITS), comprising six layers of silicon detectors with the innermost two composed of pixel detectors, and the Time Projection Chamber (TPC), a gaseous detector and the main tracking device.
%, measuring up to 159 space points per track. 
%These detectors are located inside a solenoid magnet, providing a uniform magnetic field of 0.5 T parallel to the beam axis. 
Tracks were required to have hits on both pixel layers of the ITS and a distance of closest approach to the primary vertex of less than 1 cm along the beam axis and 0.25 cm in the transverse plane, to reduce the contamination of electrons from photon conversions and particle weak decays \cite{ALICE-PUBLIC-2017-005}.
%Tracks were required to have hits on both pixel layers of the ITS to reduce the contamination of electrons from photon conversions in the detector material. In order to reject secondary electrons \cite{ALICE-PUBLIC-2017-005}, produced in interactions with the detector material or from particle weak decays, the tracks were required to have a distance of closest approach to the primary vertex of less than 1 cm along the beam axis and 0.25 cm in the transverse plane. 
The particle identification employed a selection on the specific ionisation energy loss inside the TPC of $-1 < n_\sigma^{\rm{TPC}} < 3$, where $n_\sigma$ is the difference between the measured and expected detector response signals for electrons %in a given detector, 
normalised to the response resolution.
A selection ($-3 < n _\sigma^{\rm{TOF}} < 3$) was also applied using the Time of Flight (TOF) detector to further separate hadrons and electrons.
%a set of multigap resistive plate chambers.
%that can separate hadrons and electrons at low momentum via time-of-flight measurement. 
%with an acceptance of $|\eta|<0.9$ and full azimuthal coverage.
%{\color {red} It also relies on the Time of Flight  (TOF) detector, a set of multigap resistive plate chambers that can separate hadrons and electrons at low momentum via time-of-flight measurement with an acceptance of $|\eta|<0.9$ and full azimuthal coverage, where tracks with $-3 < n _\sigma^{\rm{TOF}} < 3$ were selected. }
%information from TOF ($-3 < n _\sigma^{\rm{TOF}} < 3$), a set of multigap resistive plate chambers that can separate hadrons and electrons at low momentum via time-of-flight measurement.} 
The electron reconstruction efficiency was calculated using Monte Carlo simulations of events containing c$\bar{\rm{c}}$ and b$\bar{\rm{b}}$ pairs generated with PYTHIA 6.4.21~\cite{Sjostrand:2006za} and the Perugia-2011 tune~\cite{Skands:2009zm}, and an underlying p-Pb collision generated using HIJING 1.36~\cite{Wang:1991hta}. The generated particles were propagated through the detector using the GEANT3 transport package~\cite{GEANT3}. With the selections described above, the resulting electron reconstruction efficiency is about 28\% (32\%) at $p_{\rm T}^{\rm e} = 1.5$ GeV/$c$ (6 GeV/$c$). The contamination from charged hadrons, estimated as described in~\cite{Abelev:2012xe}, amounts to about 1\% (10\%) for $1.5 < p_{\rm T}^{\rm{e}} < 4$ GeV/$c$ ($4 < p_{\rm T}^{\rm{e}} < 6$ GeV/$c$). 

The selected electrons are composed of signal heavy-flavour decay electrons (HFe), originating from semi-leptonic decays of open heavy-flavour hadrons, and background electrons.
The main background sources are photon conversions ($\gamma \rightarrow \rm{e}^+ \rm{e}^-$) in the beam vacuum tube and in the material of the innermost ITS layers, and Dalitz decays of neutral mesons ($\pi^{0} \rightarrow \gamma ~\rm{e}^+ \rm{e}^-$ and $\eta \rightarrow \gamma ~\rm{e}^+ \rm{e}^-$), defined as non-heavy-flavour decay electrons (NonHFe) hereafter. Background contributions from other Dalitz decays or decays of kaons and J/$\psi$ mesons are negligible in the $p_{\rm T}$ range studied in the analysis~\cite{Adam:2015qda} %~\cite{Abelev:2012xe}
and were not considered.
To estimate the background contribution, di-electron pairs were defined by pairing the selected electrons with opposite-charge electron partners to form unlike-signed pairs (ULS) and calculating their invariant mass ($M_{\rm{e}^+\rm{e}^-}$). 
Partner electrons were selected applying similar but looser track quality and particle identification criteria than those used for selecting signal electrons. 
The di-electron pairs from NonHFe %the non-heavy-flavour decay electron 
sources have a small invariant mass, while heavy-flavour decay electrons can form ULS pairs mainly through random combinations with other %background 
electrons, resulting in a continuous invariant-mass distribution. 
%distribution in the low mass region.
The combinatorial contribution was estimated from the invariant mass distribution of like-signed electron (LS) pairs.
%The combinatorial contribution was estimated by calculating the invariant mass between the selected electrons and like-sign electron partners (LS). 
The NonHFe background contribution was then evaluated by subtracting the LS distribution from the ULS distribution in the invariant mass region $M_{\rm{e}^+\rm{e}^-} <$ 140 MeV/$c^2$. More details on the procedure can be found in~\cite{Abelev:2014hla,Adam:2015qda}.
%This procedure is explained in detail in previous ALICE papers~\cite{Abelev:2014hla,Adam:2015qda}. 
%{\color{red}{The di-electron pairs from the non-heavy-flavour electrons (NonHFe) sources have a small invariant mass, while heavy-flavour decay electrons do not have an ULS partner.}}
%The invariant mass distribution for these non-heavy-flavour electrons (NonHFe) is peaked at low $M_{\rm{e}^+\rm{e}^-}$, while no correlation signal is present in the low $M_{\rm{e}^+\rm{e}^-}$ region when pairing electrons from heavy-flavour decays. 
%Only the invariant mass region, $M_{\rm{e}^+\rm{e}^-} <$ 140 MeV/$c^2$, is considered for the evaluation of the background. {\color{red}{The ULS invariant mass distribution also includes random combination of ULS pairs where a small fraction of HFe were wrongly paired with a background electron, in the low mass region}}. This combinatorial background was obtained by calculating the invariant mass between the selected electrons and like-sign electron partners (LS). 
The efficiency ($\varepsilon_{\rm{NonHFe}}$) of finding the partner electron to identify non-heavy-flavour decay electrons was calculated with the aforementioned Monte Carlo simulations, %used to estimate the electron reconstruction efficiency, 
and is about 60\% for $1.5 < p_{\rm T}^{\rm{e}} < 2$ GeV/$c$, rising to 76\% for $4 < p_{\rm T}^{\rm{e}} < 6$ GeV/$c$.

The number of heavy-flavour decay electrons ($N_{\rm{HFe}}$) can be expressed as:
\begin{equation}
%\begin{split}
N_{\rm{HFe}} = N_{\rm{e}} - N_{\rm{NonHFe}} = N_{\rm{e}} - \frac{1}{\epsilon_{\rm{NonHFe}}} (N_{\rm{ULSe}} - N_{\rm{LSe }}),
\label{Eq:NHFE}
%\end{split}
\end{equation}
where $N_{\rm{ULSe}}$ and $N_{\rm{LSe}}$ are the number of electrons which form unlike-sign and like-sign pairs, respectively, with $M_{\rm{e}^+\rm{e}^-} < 140$ MeV/$c^2$, %used to obtain the number of non-heavy-flavour background electrons ($N_{\rm{NonHFe}}$), 
 and $N_{\rm{e}}$ is the number of selected electrons.
 %after subtracting the hadron contamination.
%$N_{ULS}$ and $N_{LS}$ are the number of unlike-sign and like-sign electrons used to obtain the number of non heavy-flavour background electrons ($N_{NonHFe}$). 

The two-particle correlation distributions between electrons (trigger) and charged (associated) particles  were obtained for three different $p_{\rm{T}}^{\rm{e}}$ intervals ($1.5 < p_{\rm{T}}^{\rm{e}} < 2$ GeV/$c$, $2 < p_{\rm{T}}^{\rm{e}} < 4$ GeV/$c$ and $4 < p_{\rm{T}}^{\rm{e}} < 6$ GeV/$c$). Associated charged particles with $0.3 < p_{\rm T}^{\rm{ch}} < 2$ GeV/$c$ and $|\eta| <$ 0.8 were selected with similar criteria as used for electrons, apart from requiring a hit in at least one, instead of both, of the two pixel layers and not applying any particle identification. 
The single-track reconstruction efficiency and the contamination from secondary particles \cite{ALICE-PUBLIC-2017-005} were estimated using Monte Carlo simulations of p--Pb collisions produced with the DPMJET 3.0 event generator \cite{Roesler:2000he} and GEANT3~\cite{GEANT3} for the particle transport. %performing the particle transport with GEANT3
Both were found to be independent of the event multiplicity. With the selections described above, the tracking efficiency varies from 75\% to 85\% depending on track momentum and primary vertex position, and the contamination of secondary particles varies from 3\% to 5.5\% with decreasing $p_{\rm{T}}^{\rm{ch}}$. 

The $(\Delta\varphi,\Delta\eta)$ %angular 
correlation distribution between heavy-flavour decay electrons and charged particles is obtained with the equation: %using cp-orrelation distributions obtained for each of the components on the right side of the Eq.\ref{Eq:NHFE} as:
\begin{equation}
\begin{split}
S_{\rm{HFe}} & = S_{\rm{e}} - S_{\rm{NonHFe}} \\
&  = S_{\rm{e}} - S_{\rm{NonHFe}}^{\rm{ID}} - S_{\rm{NonHFe}}^{\rm{nonID}} \\
& { = S_{\rm{e}} - S_{\rm{NonHFe}}^{\rm{ID}} - \left(\frac{1}{\varepsilon_{\rm{NonHFe}}}-1\right) S_{\rm{NonHFe}}^{\rm{ID*}} },
%%\color{red}{S} & \color{red}{ = \frac{{\rm{d}^2} N}{\rm{d}\Delta\eta \rm{d}\Delta\varphi}}
\label{eq:Sdistribution}
\end{split}
\end{equation}

where $S$ corresponds to ${{\rm{d}^2} N_{\rm{e-ch}}(\Delta\eta,\Delta\varphi)}/{\rm{d}\Delta\eta \rm{d}\Delta\varphi}$. The correlation distributions for all trigger electrons and for non-heavy-flavour decay trigger electrons are denoted as $S_{\rm{e}}$ and $S_{\rm{NonHFe}}$, respectively. The hadron contamination in $S_{\rm{e}}$ is statistically removed by subtracting a scaled di-hadron correlation distribution. The $S_{\rm{NonHFe}}$ distribution is evaluated from its two contributions $S^{\rm{ID}}_{\rm{NonHFe}}$ and $S^{\rm{nonID}}_{\rm{NonHFe}}$. The former corresponds to correlations from background electron triggers with an identified electron partner, and the latter to the expected contribution from background trigger electrons without an identified partner. The identified background distribution, $S_{\rm{NonHFe}}^{\rm{ID}}$, is evaluated using correlations of trigger electrons paired with unlike-sign and like-sign electrons, with a similar procedure as that used to evaluate $N_{\rm{NonHFe}}$ (see Eq.~\ref{Eq:NHFE}). The non-identified distribution, $S_{\rm{NonHFe}}^{\rm{nonID}}$, is estimated assuming that both identified and non-identified NonHFe %non-heavy-flavour decay electron 
triggers have the same correlation distribution, apart from reconstructed partner electrons used to calculate $M_{\rm{e}^+\rm{e}^-}$, which are removed from $S_{\rm{NonHFe}}^{\rm{ID}}$ to obtain $S_{\rm{NonHFe}}^{\rm{ID*}}$. 

The correlation distribution for heavy-flavour decay electrons was corrected for the electron and charged particle reconstruction efficiencies and for the secondary particle contamination. It was also corrected for the limited two-particle acceptance and detector inhomogeneities using the event mixing technique~\cite{ABELEV:2013wsa}. The mixed-event correlation distribution was obtained by combining electrons in an event with charged particles from other events with similar multiplicity and primary vertex position.
The correlation distribution for heavy-flavour decay electrons was divided by the number of heavy-flavour decay trigger electrons ($N_{\rm{HFe}}$, from Eq. \ref{Eq:NHFE}) corrected by their reconstruction efficiency. 

\begin{figure}
\centering
\includegraphics[width= 10cm]{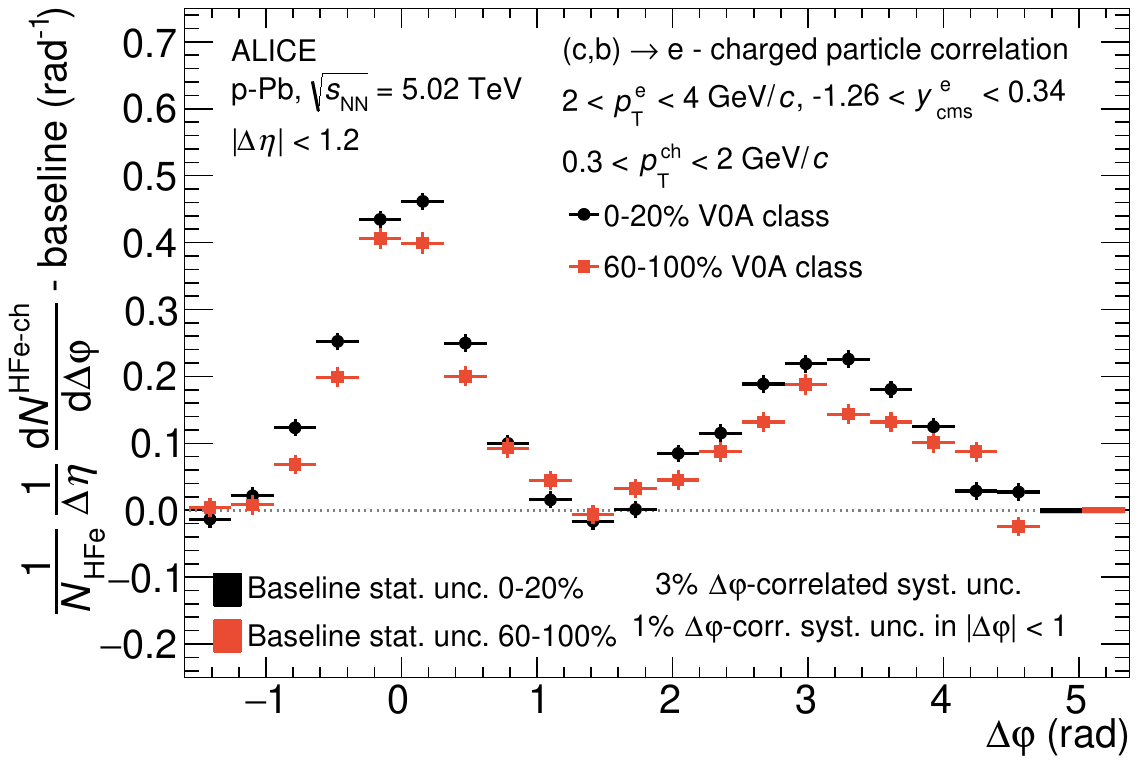}
\caption{Azimuthal correlations between heavy-flavour decay electrons and charged particles, for high-multiplicity and low-multiplicity p--Pb collisions, after subtracting the baseline (see text for details) for $2 < p_{\rm{T}}^{\rm e} < 4$ GeV/$c$ and $0.3 < p_{\rm T}^{\rm{ch}} < 2$ GeV/$c$. %Statistical uncertainties are shown as error bars. The statistical uncertainties on the baseline subtraction are represented as boxes at $\Delta\varphi \approx 5$.
}\label{fig:CorrelationFunction}
\end{figure}

\begin{figure}
\centering
\includegraphics[width= 10cm]{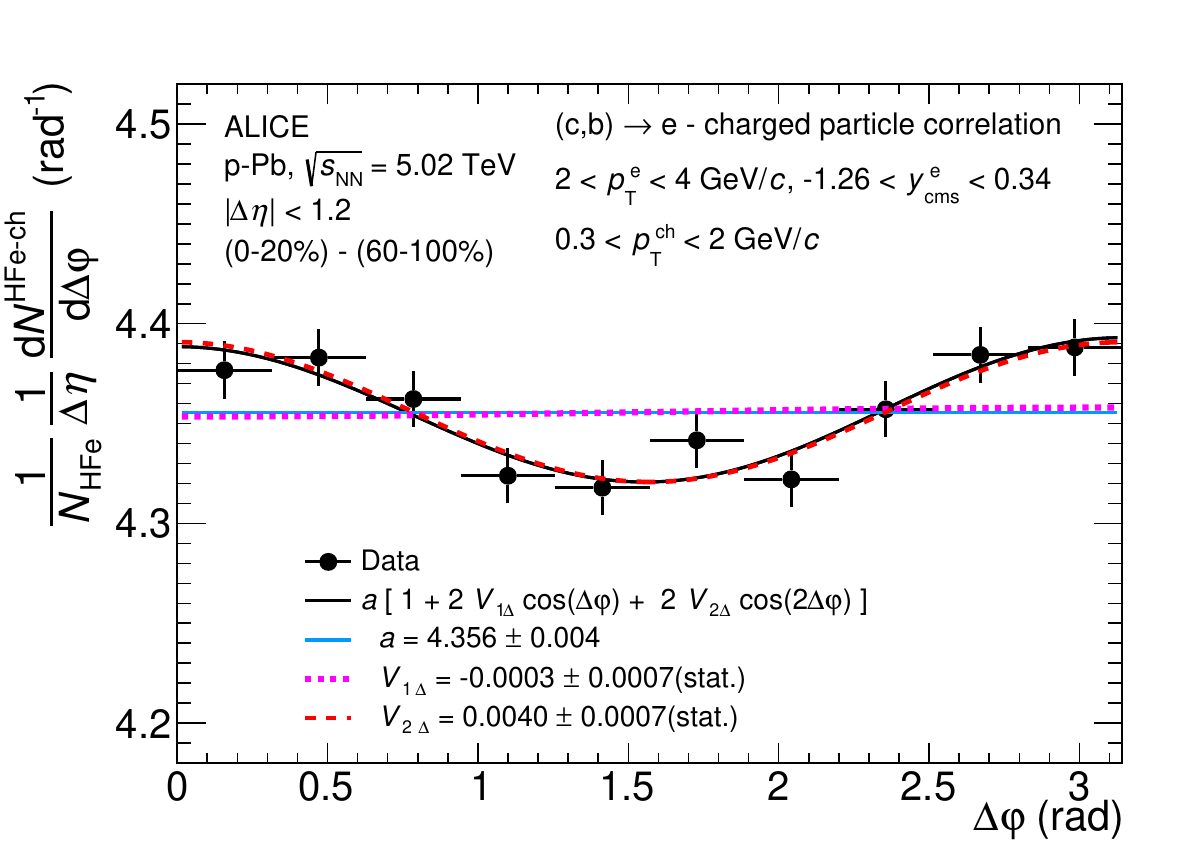}
\caption{Best fit (Eq. \ref{Eq:fit}) to the azimuthal correlation distribution between heavy-flavour decay electrons and charged particles, for high-multiplicity p--Pb collisions after subtracting the jet contribution based on low-multiplicity collisions. The distribution is shown for $2 < p_{\rm{T}}^{\rm{e}} < 4$ GeV/$c$ and $0.3 < p_{\rm{T}}^{\rm{ch}} < 2$ GeV/$c$. The figure shows only statistical uncertainty.
%The cyan solid line, pink dotted and red dashed lines indicate the fit parameter $a$, and the $\cos(\Delta\varphi)$, $\cos(2\Delta\varphi)$ modulations around that value due to the $V_{1\Delta}$ and $V_{2\Delta}$ terms, respectively.
}
\label{fig:fit}
\end{figure}

The two-dimensional correlation distribution was projected onto $\Delta \varphi$ for $|\Delta \eta| < $ 1.2 and divided by the width of the selected $\Delta\eta$ interval.
%In order to compare the jet-induced peaks from different multiplicity ranges, 
A baseline term, constant in $\Delta\varphi$, was subtracted from the correlation distributions. Its values, reported in Table \ref{table:results}, were calculated as the weighted average of the three lowest points of the distribution, following the ``zero yield at minimum'' approach~\cite{Adler:2005ee}.
%{\color{red} The baseline values are $4.312\pm 0.008$(stat)$\pm0.13$(syst), $4.3301\pm0.007$(stat)$\pm0.13$(syst) and $4.754\pm0.020$(stat)$\pm0.14$(syst) for high multiplicity collisions; and $1.235\pm0.006$(stat)$\pm0.037$(syst), $1.294\pm0.008$(stat)$+-0.038$(syst) and $1.433\pm0.022$(stat)$\pm0.043$(syst) for low multiplicity collisions,  both in the electron $p_{\rm T}$ intervals $1.5 < p_{\rm{T}}^{\rm e}<2$ GeV/$c$, $2 < p_{\rm{T}}^{\rm{e}} <4$ GeV/$c$ and $4 < p_{\rm{T}}^{\rm{e}} <6$ GeV/$c$, respectively.}
%,  and $ for high multiplicity collisions; and 
%$ and(syst)
%The values of $V_{2\Delta}^{\rm{HFe-ch}}$ obtained from the fit in the three $p_{\rm T}^{\rm e}$ intervals are  and $0.0019 \pm 0.0019$(stat) $\pm$ 0.0003(syst) for $1.5 < p_{\rm{T}}^{\rm e}<2$ GeV/$c$, $2 < p_{\rm{T}}^{\rm{e}} <4$ GeV/$c$ and $4 < p_{\rm{T}}^{\rm{e}} <6$ GeV/$c$,
The resulting correlation distributions in the two considered multiplicity classes (0--20\% and 60--100\%) are shown in Fig.~\ref{fig:CorrelationFunction} for $2 < p_{\rm{T}}^{\rm e} < 4$ GeV/$c$. 
An enhancement of the near- and away-side peaks is present in high-multiplicity collisions. To study this feature, the baseline-subtracted correlation distribution obtained in low-multiplicity events was subtracted from the correlation distribution measured in high-multiplicity events, as described in~\cite{Abelev:2012ola}. 
This removes the jet-induced correlation peaks, assuming that they are the same in low- and high-multiplicity events. 
The correlation distribution was restricted to the (0,$\pi$) range by reflecting the symmetrical points. The resulting distribution, shown in Fig.~\ref{fig:fit}, was fitted with the Fourier decomposition of Eq. \ref{Eq:fit}. An azimuthal anisotropy, dominated by the second-order term $V_{2\Delta}^{\rm HFe-ch}$, was found.
\begin{equation}
\frac{1}{\Delta\eta} \frac{1}{N_{\rm{HFe}}} \frac{{\rm{d}}N_{\rm{HFe-ch}}(\Delta\varphi)}{\rm{d}\Delta\varphi} = a [1+ 2 V_{1\Delta}^{\rm{HFe-ch}}\cos(\Delta\varphi) + 2 V_{2\Delta}^{\rm{HFe-ch}}\cos(2\Delta\varphi)]
\label{Eq:fit}
\end{equation}

The measured $V_{2\Delta}^{\rm HFe-ch}$ in high-multiplicity events does not exclude the possibility of having a $V_{2\Delta}^{\rm HFe-ch}$ contribution in the low-multiplicity events, as described in~\cite{Aaboud:2016yar}.

\begin{table*}[b]
\centering
\begin{tabular}{c|ccc}
$p_{\rm T}^{\rm e}$ (GeV/$c$) & $V_{2\Delta}^{\rm{HFe-ch}}$ $\pm$ stat $\pm$ syst  & $b_{\rm LM}$ $\pm$ stat $\pm$ syst ($\text{rad}^{-1}$) & $b_{\rm HM}$ $\pm$ stat $\pm$ syst ($\text{rad}^{-1}$) \\ \hline
$1.5< p_{\rm T}^e < 2$ & $ (38 \pm 8\pm 6) \times 10^{-4}$ & $1.235\pm0.006\pm0.037$ & $4.312\pm 0.008\pm0.129$ \\ 
$2< p_{\rm T}^e < 4$ & $(40 \pm 7\pm 5)\times 10^{-4}$& $1.294\pm0.008\pm0.038$ & $4.330\pm 0.007\pm0.129$ \\ 
$4< p_{\rm T}^e < 6$ & $(19 \pm 19\pm 3)\times 10^{-4}$&  $1.433\pm0.022\pm0.043$ & $4.754\pm0.020\pm0.142$ \\
\end{tabular}
\caption{Results for $V_{2\Delta}^{\rm{HFe-ch}}$ and baselines in high- ($b_{\rm HM}$) and low-multiplicity ($b_{\rm LM}$) collisions.}\label{table:results}
\end{table*}

The systematic uncertainties on the azimuthal correlation distribution can originate from: (i) potential biases in the procedure employed to select electron candidates and estimate the hadron contamination, (ii) %in the 
removal of the background electrons not produced in heavy-flavour hadron decays and (iii) %in the 
choice of the associated particle selection.
%The systematic uncertainties on the correlation distribution from the biases introduced by the procedure employed to select electron candidates and estimate the hadron contamination from the electron sample, removal of the non-heavy flavour background electrons and selection criteria of the associated particles are studied. 
A systematic uncertainty related to the electron reconstruction efficiency arises from imprecisions in the description of the detector response. It was studied by varying the electron selection in the ITS and TPC. The uncertainty affecting the removal of the hadron contamination was estimated by varying the particle identification criteria in the TPC ($n_\sigma^{\rm{TPC}}$). A total uncertainty of less than 0.5\% was estimated from these sources. The uncertainty related to the efficiency of finding the partner electron and to the stability of the $S_{\rm{NonHFe}}$ distribution 
%, evaluated from its two contributions $S^{\rm{ID}}_{\rm{NonHFe}}$ and $S^{\rm{nonID}}_{\rm{NonHFe}}$, 
was studied by varying the selection for partner tracks and pair invariant mass, resulting in an uncertainty of less than 0.5\%. The uncertainty on the associated track reconstruction efficiency, obtained by varying the associated track selection criteria and by comparing the probabilities of track prolongation from TPC to ITS in data and simulations, was estimated to be 3\%~\cite{ALICE:2012mj}. A systematic effect due to the contamination of the associated particles by secondaries comes from residual discrepancy between Monte Carlo and data in the relative abundances of particle species and 
%a possible bias in the abundances of particle species in the simulation and 
was studied by varying the selection on the distance of closest approach to the primary vertex. It was quantified to be 1\% (correlated in $\Delta \varphi$), with an additional 1\% (correlated) for $|\Delta\varphi| < $1. Combining the uncertainties from all the above sources results in a 3\% total systematic uncertainty (correlated in $\Delta \varphi$) and an additional 1\% (also correlated) for $|\Delta \varphi| < 1$.

%The systematic uncertainties from all the sources described in the previous paragraph were also studied for the $V_{2\Delta}$ measurement. 
The systematic uncertainties from the above mentioned sources are also present in the $V_{2\Delta}^{\rm HFe-ch}$. 
The uncertainty related to the electron selection and the identification of non-heavy-flavour decay electrons on $V_{2\Delta}^{\rm HFe-ch}$ were quantified to be about 2--3\% and 5\%, respectively. The contamination of the associated particles by secondaries leads to a 3\% systematic uncertainty.
In order to test whether the observed $\Delta\varphi$ modulation and the non-zero $V_{2\Delta}^{\rm HFe-ch}$ could originate from a residual jet contribution, due to possible differences between the jet structures in low- and high-multiplicity collisions, the $\Delta\eta$ integration region was modified by excluding central intervals of $|\Delta\eta| < \Delta\eta^{gap}$, varying $\Delta\eta^{gap}$ from 0.2 to 0.6.
%In order to test whether the observed modulation and the non-zero $V_{2\Delta}^{\rm HFe-ch}$ could originate from a residual jet contribution, due to possible differences between the jet structures in low- and high-multiplicity collisions,
%%derive from differences in the jet-related correlation terms in high- and low-multiplicity collisions, 
%the $\Delta\eta$ range used to obtain the $\Delta\varphi$ projection was varied by introducing pseudorapidity gaps from 0.2 to 0.6 units. 
The observed variation on $V_{2\Delta}^{\rm HFe-ch}$ was 11--15\%, depending on the electron $p_{\rm T}$ interval, and was taken as the systematic uncertainty from the jet subtraction. The stability of the $V_{2\Delta}^{\rm HFe-ch}$ value against the variation of the $\Delta\eta$ range suggests a long-range nature of the observed anisotropy. The inclusion of a $V_{3\Delta}^{\rm{HFe-ch}}$ term in the fit function, in Eq.~\ref{Eq:fit}, affects the $V_{2\Delta}^{\rm HFe-ch}$ estimation by less than 0.5\%.
Combining the different uncertainty sources results in a total systematic uncertainty on $V_{2\Delta}^{\rm HFe-ch}$ of 13--16\% depending on $p_{\rm T}^{\rm e}$.

The values of $V_{2\Delta}^{\rm{HFe-ch}}$ obtained from the fits are reported in Table \ref{table:results}.
%in the three $p_{\rm T}^{\rm e}$ intervals are $0.0038 \pm 0.0008$(stat) $\pm$ 0.0006(syst), $0.0040 \pm 0.0007$(stat) $\pm$ 0.0005(syst) and $0.0019 \pm 0.0019$(stat) $\pm$ 0.0003(syst) for $1.5 < p_{\rm{T}}^{\rm e}<2$ GeV/$c$, $2 < p_{\rm{T}}^{\rm{e}} <4$ GeV/$c$ and $4 < p_{\rm{T}}^{\rm{e}} <6$ GeV/$c$, respectively. 
%The $V_{1\Delta}^{\rm{HFe-ch}}$ fit values are compatible with zero in all the $p_{\rm T}^{\rm e}$ intervals. 
The measured $V_{2\Delta}^{\rm{HFe-ch}}$ is larger than zero with a significance of 4.6$\sigma$ for the $2 < p_{\rm{T}}^{\rm{e}} <4$ GeV/$c$ range.
The significance for $V_{2\Delta}^{\rm{HFe-ch}} > 0$ in the 
%combined $p_{\rm T}^{\rm e}$ 
interval $1.5 < p_{\rm{T}}^{\rm{e}} <4$ GeV/$c$, considering statistical and systematic uncertainties, is about 6$\sigma$.
%The significance for $V_{2\Delta}^{\rm{HFe-ch}} > 0$ in at least one of the $p_{\rm T}^{\rm e}$ intervals, $1.5 < p_{\rm{T}}^{\rm{e}} <2$ GeV/$c$ and $2 < p_{\rm{T}}^{\rm{e}} <4$ GeV/$c$, combining statistical and systematical uncertainties, is about 6$\sigma$.

\begin{figure}
\centering
\includegraphics[width= 10cm]{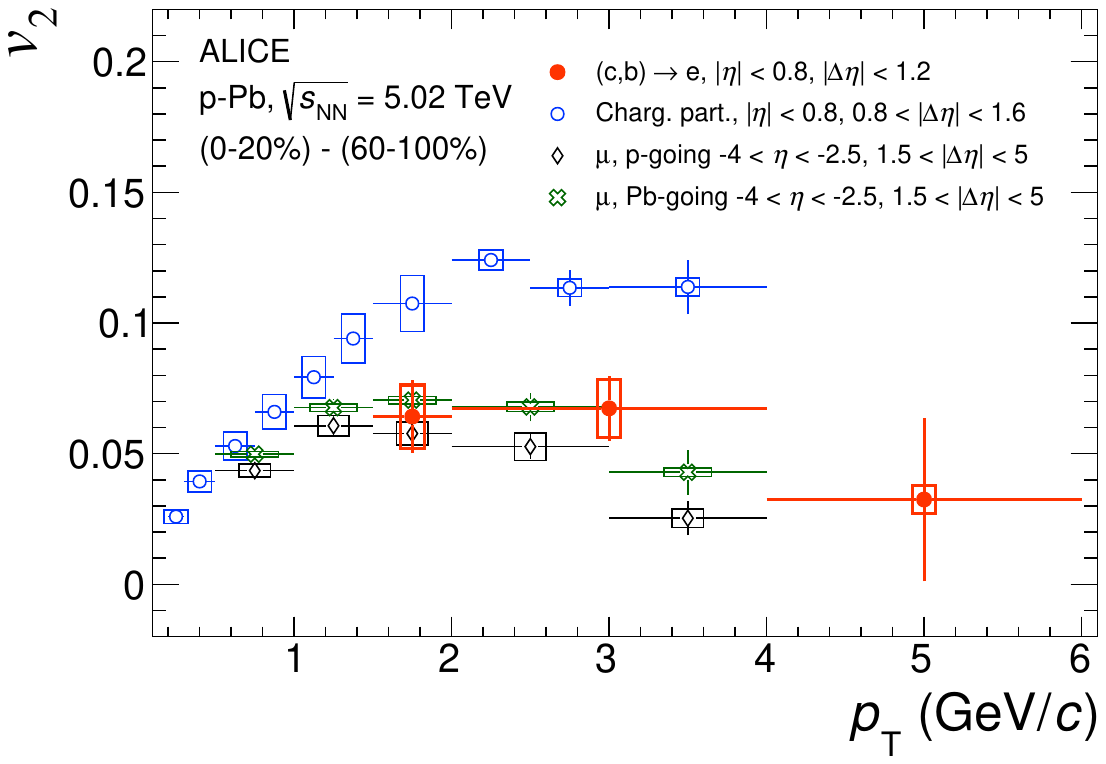}
\caption{Heavy-flavour decay electron $v_2$ as a function of transverse momentum compared to the $v_2$ of unidentified charged particles~\cite{ABELEV:2013wsa} and inclusive muons~\cite{Adam:2015bka}. 
%Statistical and systematic uncertainties are shown as bars and boxes, respectively.
} %I don't think we need to say that they are 'extracted via two-particle correlations after subtraction of the jet contribution', but if you want we can add it.
\label{fig:v2Results}
\end{figure}

% The $V_{2\Delta}^{\rm{HFe-ch}}$ can be factorised [cite] as a product of heavy-flavour decay electron ($v_2^{\rm{HFe}}$) and charged particles ($v_2^{\rm{ch}}$) second-order Fourier coefficients of the respective azimuthal distributions, as $v_2^{\rm{HFe}}$ = $V_{2\Delta}^{\rm{HFe-ch}}/v_2^{\rm{ch}}$. 
Assuming its factorization in single-particle $v_{2}$ coefficients~\cite{ABELEV:2013wsa}, the $V_{2\Delta}^{\rm{HFe-ch}}$ can be expressed as the product of the second-order Fourier coefficients of the heavy-flavour decay electron ($v_2^{\rm{HFe}}$) and charged particle ($v_2^{\rm{ch}}$) azimuthal distributions, hence $v_2^{\rm{HFe}}$ = $V_{2\Delta}^{\rm{HFe-ch}}/v_2^{\rm{ch}}$. The $v_2^{\rm{ch}}$ value in the range $0.3< p_{\rm{T}}^{\rm{ch}} <2$ GeV/$c$ was obtained from the weighted average of the values measured in smaller $p_{\rm{T}}^{\rm{ch}}$ ranges in~\cite{ABELEV:2013wsa}, providing $v_2^{\rm{ch}}$ = 0.0594 $\pm$ 0.0010(stat) $\pm$ 0.0059 (syst).
The obtained $v_{2}^{\rm{HFe}}$ values are reported in Fig.~\ref{fig:v2Results} and compared to $v_{2}$ of charged particles, dominated by light-flavour hadrons, and to inclusive muons at large rapidity, mostly originating from heavy-flavour hadron decays for $p_{\rm{T}}^{\mu} > 2$ GeV/$c$. 
 The heavy-flavour decay electron $v_{2}$ is lower than $v_2^{\rm{ch}}$, though the uncertainties are large and the $p_{\rm{T}}$ interval of electron parents (heavy-flavour hadrons) is considerably broader than the range addressed in the light-flavour hadron measurement.
%$v_{2}^{\rm{HFe}}$ shows hints of being lower, though consistent within uncertainties, compared to charged-particle particle $v_{2}$. Anyway, the $p_{\rm{T}}$ interval of electron parents (heavy-flavour hadrons) is considerably broader than the range addressed in the light-flavour hadron measurement. 
The $v_2$ values for heavy-flavour electrons and inclusive muons are similar, although a direct comparison is not straightforward, given the different rapidities and the contamination in the muon sample for $p_{\rm{T}}^{\mu} < 2$ GeV/$c$. 
%The comparison of $v_2^{\rm HFe}$ at mid-rapidity with $v_{2}$ of inclusive muons at forward and backward rapidity is not straightforward, due to the different cold nuclear matter effects affecting heavy-flavour production at different rapidities \cite{Acharya:2017hdv} and to the non-heavy-flavour contamination for muons at low $p_{\rm T}^{\rm \mu}$. A comparison of $v_2^{\rm{HFe}}$ with the J/$\psi$ results \cite{Acharya:2017tfn} is also challenging, considering the different fragmentation process of heavy quarks to open and hidden mesons, and is not presented here. 
The $v_2^{\rm HFe}$ in p--Pb collisions has similar magnitude as the one in non-central Pb--Pb collisions at $\sqrt{s_{\rm NN}} = 2.76$ TeV~\cite{Adam:2016ssk}. The significance for $v_{2}^{\rm{HFe}} > 0$ is 5.1$\sigma$ for $1.5 < p_{\rm{T}}^{\rm{e}} < 4$ GeV/$c$, providing very strong indications for the presence of long-range anisotropies for heavy-flavour particles in high-multiplicity p--Pb collisions. 

%% file: Conclusion.tex
In summary, we report the measurement of $v_{2}$ for open heavy-flavour particles at mid-rapidity in high-multiplicity p--Pb collisions. The analysis was carried out via a Fourier decomposition of the azimuthal correlation distribution between heavy-flavour decay electrons and charged particles. After removing the jet contribution and fitting the high-multiplicity correlation distributions, a $V_{2\Delta}$-like modulation was obtained, qualitatively similar to the one observed for charged particles \cite{Abelev:2012ola}. The measured heavy-flavour decay electron $v_{2}$ is positive with a significance of more than 5$\sigma$ in the $1.5 < p_{\rm T}^{\rm e} < 4$ GeV/$c$ range. Its values are possibly lower %, though still compatible, 
than charged-particle $v_2$~\cite{Abelev:2012ola}, and similar to inclusive muon $v_2$ at large rapidity~\cite{Adam:2015bka}. Complementing previous results for light-flavour hadrons~\cite{Abelev:2012ola}, this measurement provides new information on the behaviour of heavy-flavour hadrons to understand the azimuthal anisotropies observed in small collision systems.

%% file: fa_2018-05-09.tex
% Version: 2018-05-09

The ALICE Collaboration would like to thank all its engineers and technicians for their invaluable contributions to the construction of the experiment and the CERN accelerator teams for the outstanding performance of the LHC complex.
The ALICE Collaboration gratefully acknowledges the resources and support provided by all Grid centres and the Worldwide LHC Computing Grid (WLCG) collaboration.
The ALICE Collaboration acknowledges the following funding agencies for their support in building and running the ALICE detector:
A. I. Alikhanyan National Science Laboratory (Yerevan Physics Institute) Foundation (ANSL), State Committee of Science and World Federation of Scientists (WFS), Armenia;
Austrian Academy of Sciences and Nationalstiftung f\"{u}r Forschung, Technologie und Entwicklung, Austria;
Ministry of Communications and High Technologies, National Nuclear Research Center, Azerbaijan;
Conselho Nacional de Desenvolvimento Cient\'{\i}fico e Tecnol\'{o}gico (CNPq), Universidade Federal do Rio Grande do Sul (UFRGS), Financiadora de Estudos e Projetos (Finep) and Funda\c{c}\~{a}o de Amparo \`{a} Pesquisa do Estado de S\~{a}o Paulo (FAPESP), Brazil;
Ministry of Science \& Technology of China (MSTC), National Natural Science Foundation of China (NSFC) and Ministry of Education of China (MOEC) , China;
Ministry of Science and Education, Croatia;
Ministry of Education, Youth and Sports of the Czech Republic, Czech Republic;
The Danish Council for Independent Research | Natural Sciences, the Carlsberg Foundation and Danish National Research Foundation (DNRF), Denmark;
Helsinki Institute of Physics (HIP), Finland;
Commissariat \`{a} l'Energie Atomique (CEA) and Institut National de Physique Nucl\'{e}aire et de Physique des Particules (IN2P3) and Centre National de la Recherche Scientifique (CNRS), France;
Bundesministerium f\"{u}r Bildung, Wissenschaft, Forschung und Technologie (BMBF) and GSI Helmholtzzentrum f\"{u}r Schwerionenforschung GmbH, Germany;
General Secretariat for Research and Technology, Ministry of Education, Research and Religions, Greece;
National Research, Development and Innovation Office, Hungary;
Department of Atomic Energy Government of India (DAE), Department of Science and Technology, Government of India (DST), University Grants Commission, Government of India (UGC) and Council of Scientific and Industrial Research (CSIR), India;
Indonesian Institute of Science, Indonesia;
Centro Fermi - Museo Storico della Fisica e Centro Studi e Ricerche Enrico Fermi and Istituto Nazionale di Fisica Nucleare (INFN), Italy;
Institute for Innovative Science and Technology , Nagasaki Institute of Applied Science (IIST), Japan Society for the Promotion of Science (JSPS) KAKENHI and Japanese Ministry of Education, Culture, Sports, Science and Technology (MEXT), Japan;
Consejo Nacional de Ciencia (CONACYT) y Tecnolog\'{i}a, through Fondo de Cooperaci\'{o}n Internacional en Ciencia y Tecnolog\'{i}a (FONCICYT) and Direcci\'{o}n General de Asuntos del Personal Academico (DGAPA), Mexico;
Nederlandse Organisatie voor Wetenschappelijk Onderzoek (NWO), Netherlands;
The Research Council of Norway, Norway;
Commission on Science and Technology for Sustainable Development in the South (COMSATS), Pakistan;
Pontificia Universidad Cat\'{o}lica del Per\'{u}, Peru;
Ministry of Science and Higher Education and National Science Centre, Poland;
Korea Institute of Science and Technology Information and National Research Foundation of Korea (NRF), Republic of Korea;
Ministry of Education and Scientific Research, Institute of Atomic Physics and Romanian National Agency for Science, Technology and Innovation, Romania;
Joint Institute for Nuclear Research (JINR), Ministry of Education and Science of the Russian Federation and National Research Centre Kurchatov Institute, Russia;
Ministry of Education, Science, Research and Sport of the Slovak Republic, Slovakia;
National Research Foundation of South Africa, South Africa;
Centro de Aplicaciones Tecnol\'{o}gicas y Desarrollo Nuclear (CEADEN), Cubaenerg\'{\i}a, Cuba and Centro de Investigaciones Energ\'{e}ticas, Medioambientales y Tecnol\'{o}gicas (CIEMAT), Spain;
Swedish Research Council (VR) and Knut \& Alice Wallenberg Foundation (KAW), Sweden;
European Organization for Nuclear Research, Switzerland;
National Science and Technology Development Agency (NSDTA), Suranaree University of Technology (SUT) and Office of the Higher Education Commission under NRU project of Thailand, Thailand;
Turkish Atomic Energy Agency (TAEK), Turkey;
National Academy of  Sciences of Ukraine, Ukraine;
Science and Technology Facilities Council (STFC), United Kingdom;
National Science Foundation of the United States of America (NSF) and United States Department of Energy, Office of Nuclear Physics (DOE NP), United States of America.

%% file: Alice_Authorlist_2018-May-01.tex
% Collaboration: CERN-LHC-ALICE
% Generation Date is 2018-May-01

% How to use:
%%%%%%%%% appendix with author list
%\appendix
%\section{The ALICE Collaboration}
%\label{app:collab}
%\input{Alice_Authorslist_XXXX-Axx-XX.tex}
\begingroup
\small
\begin{flushleft}
S.~Acharya\Irefn{org139}\And 
F.~Torales-Acosta\Irefn{org20}\And 
D.~Adamov\'{a}\Irefn{org93}\And 
J.~Adolfsson\Irefn{org80}\And 
M.M.~Aggarwal\Irefn{org98}\And 
G.~Aglieri Rinella\Irefn{org34}\And 
M.~Agnello\Irefn{org31}\And 
N.~Agrawal\Irefn{org48}\And 
Z.~Ahammed\Irefn{org139}\And 
S.U.~Ahn\Irefn{org76}
% ...
\And
\renewcommand\labelenumi{\textsuperscript{\theenumi}~}

\section*{Affiliation notes}
\renewcommand\theenumi{\roman{enumi}}
\begin{Authlist}
\item \Adef{org*}Deceased
\item \Adef{orgI}Dipartimento DET del Politecnico di Torino, Turin, Italy
\item \Adef{orgII}M.V. Lomonosov Moscow State University, D.V. Skobeltsyn Institute of Nuclear, Physics, Moscow, Russia
\item \Adef{orgIII}Department of Applied Physics, Aligarh Muslim University, Aligarh, India
\item \Adef{orgIV}Institute of Theoretical Physics, University of Wroclaw, Poland
\end{Authlist}

\section*{Collaboration Institutes}
\renewcommand\theenumi{\arabic{enumi}~}
\begin{Authlist}
\item \Idef{org1}A.I. Alikhanyan National Science Laboratory (Yerevan Physics Institute) Foundation, Yerevan, Armenia
\item \Idef{org2}Bogolyubov Institute for Theoretical Physics, National Academy of Sciences of Ukraine, Kiev, Ukraine
\item \Idef{org3}Bose Institute, Department of Physics  and Centre for Astroparticle Physics and Space Science (CAPSS), Kolkata, India
\item \Idef{org4}Budker Institute for Nuclear Physics, Novosibirsk, Russia
\item \Idef{org5}California Polytechnic State University, San Luis Obispo, California, United States
\item \Idef{org6}Central China Normal University, Wuhan, China
\item \Idef{org7}Centre de Calcul de l'IN2P3, Villeurbanne, Lyon, France
\item \Idef{org8}Centro de Aplicaciones Tecnol\'{o}gicas y Desarrollo Nuclear (CEADEN), Havana, Cuba
\item \Idef{org9}Centro de Investigaci\'{o}n y de Estudios Avanzados (CINVESTAV), Mexico City and M\'{e}rida, Mexico
\item \Idef{org10}Centro Fermi - Museo Storico della Fisica e Centro Studi e Ricerche ``Enrico Fermi', Rome, Italy
\item \Idef{org11}Chicago State University, Chicago, Illinois, United States
\item \Idef{org12}China Institute of Atomic Energy, Beijing, China
\item \Idef{org13}Chonbuk National University, Jeonju, Republic of Korea
\item \Idef{org14}Comenius University Bratislava, Faculty of Mathematics, Physics and Informatics, Bratislava, Slovakia
\item \Idef{org15}COMSATS Institute of Information Technology (CIIT), Islamabad, Pakistan
\item \Idef{org16}Creighton University, Omaha, Nebraska, United States
\item \Idef{org17}Department of Physics, Aligarh Muslim University, Aligarh, India
\item \Idef{org18}Department of Physics, Pusan National University, Pusan, Republic of Korea
\item \Idef{org19}Department of Physics, Sejong University, Seoul, Republic of Korea
\item \Idef{org20}Department of Physics, University of California, Berkeley, California, United States
\item \Idef{org21}Department of Physics, University of Oslo, Oslo, Norway
\item \Idef{org22}Department of Physics and Technology, University of Bergen, Bergen, Norway
\item \Idef{org23}Dipartimento di Fisica dell'Universit\`{a} 'La Sapienza' and Sezione INFN, Rome, Italy
\item \Idef{org24}Dipartimento di Fisica dell'Universit\`{a} and Sezione INFN, Cagliari, Italy
\item \Idef{org25}Dipartimento di Fisica dell'Universit\`{a} and Sezione INFN, Trieste, Italy
\item \Idef{org26}Dipartimento di Fisica dell'Universit\`{a} and Sezione INFN, Turin, Italy
\item \Idef{org27}Dipartimento di Fisica e Astronomia dell'Universit\`{a} and Sezione INFN, Bologna, Italy
\item \Idef{org28}Dipartimento di Fisica e Astronomia dell'Universit\`{a} and Sezione INFN, Catania, Italy
\item \Idef{org29}Dipartimento di Fisica e Astronomia dell'Universit\`{a} and Sezione INFN, Padova, Italy
\item \Idef{org30}Dipartimento di Fisica `E.R.~Caianiello' dell'Universit\`{a} and Gruppo Collegato INFN, Salerno, Italy
\item \Idef{org31}Dipartimento DISAT del Politecnico and Sezione INFN, Turin, Italy
\item \Idef{org32}Dipartimento di Scienze e Innovazione Tecnologica dell'Universit\`{a} del Piemonte Orientale and INFN Sezione di Torino, Alessandria, Italy
\item \Idef{org33}Dipartimento Interateneo di Fisica `M.~Merlin' and Sezione INFN, Bari, Italy
\item \Idef{org34}European Organization for Nuclear Research (CERN), Geneva, Switzerland
\item \Idef{org35}Faculty of Electrical Engineering, Mechanical Engineering and Naval Architecture, University of Split, Split, Croatia
\item \Idef{org36}Faculty of Engineering and Science, Western Norway University of Applied Sciences, Bergen, Norway
\item \Idef{org37}Faculty of Nuclear Sciences and Physical Engineering, Czech Technical University in Prague, Prague, Czech Republic
\item \Idef{org38}Faculty of Science, P.J.~\v{S}af\'{a}rik University, Ko\v{s}ice, Slovakia
\item \Idef{org39}Frankfurt Institute for Advanced Studies, Johann Wolfgang Goethe-Universit\"{a}t Frankfurt, Frankfurt, Germany
\item \Idef{org40}Gangneung-Wonju National University, Gangneung, Republic of Korea
\item \Idef{org41}Gauhati University, Department of Physics, Guwahati, India
\item \Idef{org42}Helmholtz-Institut f\"{u}r Strahlen- und Kernphysik, Rheinische Friedrich-Wilhelms-Universit\"{a}t Bonn, Bonn, Germany
\item \Idef{org43}Helsinki Institute of Physics (HIP), Helsinki, Finland
\item \Idef{org44}High Energy Physics Group,  Universidad Aut\'{o}noma de Puebla, Puebla, Mexico
\item \Idef{org45}Hiroshima University, Hiroshima, Japan
\item \Idef{org46}Hochschule Worms, Zentrum  f\"{u}r Technologietransfer und Telekommunikation (ZTT), Worms, Germany
\item \Idef{org47}Horia Hulubei National Institute of Physics and Nuclear Engineering, Bucharest, Romania
\item \Idef{org48}Indian Institute of Technology Bombay (IIT), Mumbai, India
\item \Idef{org49}Indian Institute of Technology Indore, Indore, India
\item \Idef{org50}Indonesian Institute of Sciences, Jakarta, Indonesia
\item \Idef{org51}INFN, Laboratori Nazionali di Frascati, Frascati, Italy
\item \Idef{org52}INFN, Sezione di Bari, Bari, Italy
\item \Idef{org53}INFN, Sezione di Bologna, Bologna, Italy
\item \Idef{org54}INFN, Sezione di Cagliari, Cagliari, Italy
\item \Idef{org55}INFN, Sezione di Catania, Catania, Italy
\item \Idef{org56}INFN, Sezione di Padova, Padova, Italy
\item \Idef{org57}INFN, Sezione di Roma, Rome, Italy
\item \Idef{org58}INFN, Sezione di Torino, Turin, Italy
\item \Idef{org59}INFN, Sezione di Trieste, Trieste, Italy
\item \Idef{org60}Inha University, Incheon, Republic of Korea
\item \Idef{org61}Institut de Physique Nucl\'{e}aire d'Orsay (IPNO), Institut National de Physique Nucl\'{e}aire et de Physique des Particules (IN2P3/CNRS), Universit\'{e} de Paris-Sud, Universit\'{e} Paris-Saclay, Orsay, France
\item \Idef{org62}Institute for Nuclear Research, Academy of Sciences, Moscow, Russia
\item \Idef{org63}Institute for Subatomic Physics, Utrecht University/Nikhef, Utrecht, Netherlands
\item \Idef{org64}Institute for Theoretical and Experimental Physics, Moscow, Russia
\item \Idef{org65}Institute of Experimental Physics, Slovak Academy of Sciences, Ko\v{s}ice, Slovakia
\item \Idef{org66}Institute of Physics, Homi Bhabha National Institute, Bhubaneswar, India
\item \Idef{org67}Institute of Physics of the Czech Academy of Sciences, Prague, Czech Republic
\item \Idef{org68}Institute of Space Science (ISS), Bucharest, Romania
\item \Idef{org69}Institut f\"{u}r Kernphysik, Johann Wolfgang Goethe-Universit\"{a}t Frankfurt, Frankfurt, Germany
\item \Idef{org70}Instituto de Ciencias Nucleares, Universidad Nacional Aut\'{o}noma de M\'{e}xico, Mexico City, Mexico
\item \Idef{org71}Instituto de F\'{i}sica, Universidade Federal do Rio Grande do Sul (UFRGS), Porto Alegre, Brazil
\item \Idef{org72}Instituto de F\'{\i}sica, Universidad Nacional Aut\'{o}noma de M\'{e}xico, Mexico City, Mexico
\item \Idef{org73}iThemba LABS, National Research Foundation, Somerset West, South Africa
\item \Idef{org74}Johann-Wolfgang-Goethe Universit\"{a}t Frankfurt Institut f\"{u}r Informatik, Fachbereich Informatik und Mathematik, Frankfurt, Germany
\item \Idef{org75}Joint Institute for Nuclear Research (JINR), Dubna, Russia
\item \Idef{org76}Korea Institute of Science and Technology Information, Daejeon, Republic of Korea
\item \Idef{org77}KTO Karatay University, Konya, Turkey
\item \Idef{org78}Laboratoire de Physique Subatomique et de Cosmologie, Universit\'{e} Grenoble-Alpes, CNRS-IN2P3, Grenoble, France
\item \Idef{org79}Lawrence Berkeley National Laboratory, Berkeley, California, United States
\item \Idef{org80}Lund University Department of Physics, Division of Particle Physics, Lund, Sweden
\item \Idef{org81}Nagasaki Institute of Applied Science, Nagasaki, Japan
\item \Idef{org82}Nara Women{'}s University (NWU), Nara, Japan
\item \Idef{org83}National and Kapodistrian University of Athens, School of Science, Department of Physics , Athens, Greece
\item \Idef{org84}National Centre for Nuclear Research, Warsaw, Poland
\item \Idef{org85}National Institute of Science Education and Research, Homi Bhabha National Institute, Jatni, India
\item \Idef{org86}National Nuclear Research Center, Baku, Azerbaijan
\item \Idef{org87}National Research Centre Kurchatov Institute, Moscow, Russia
\item \Idef{org88}Niels Bohr Institute, University of Copenhagen, Copenhagen, Denmark
\item \Idef{org89}Nikhef, National institute for subatomic physics, Amsterdam, Netherlands
\item \Idef{org90}NRC Kurchatov Institute IHEP, Protvino, Russia
\item \Idef{org91}NRNU Moscow Engineering Physics Institute, Moscow, Russia
\item \Idef{org92}Nuclear Physics Group, STFC Daresbury Laboratory, Daresbury, United Kingdom
\item \Idef{org93}Nuclear Physics Institute of the Czech Academy of Sciences, \v{R}e\v{z} u Prahy, Czech Republic
\item \Idef{org94}Oak Ridge National Laboratory, Oak Ridge, Tennessee, United States
\item \Idef{org95}Ohio State University, Columbus, Ohio, United States
\item \Idef{org96}Petersburg Nuclear Physics Institute, Gatchina, Russia
\item \Idef{org97}Physics department, Faculty of science, University of Zagreb, Zagreb, Croatia
\item \Idef{org98}Physics Department, Panjab University, Chandigarh, India
\item \Idef{org99}Physics Department, University of Jammu, Jammu, India
\item \Idef{org100}Physics Department, University of Rajasthan, Jaipur, India
\item \Idef{org101}Physikalisches Institut, Eberhard-Karls-Universit\"{a}t T\"{u}bingen, T\"{u}bingen, Germany
\item \Idef{org102}Physikalisches Institut, Ruprecht-Karls-Universit\"{a}t Heidelberg, Heidelberg, Germany
\item \Idef{org103}Physik Department, Technische Universit\"{a}t M\"{u}nchen, Munich, Germany
\item \Idef{org104}Research Division and ExtreMe Matter Institute EMMI, GSI Helmholtzzentrum f\"ur Schwerionenforschung GmbH, Darmstadt, Germany
\item \Idef{org105}Rudjer Bo\v{s}kovi\'{c} Institute, Zagreb, Croatia
\item \Idef{org106}Russian Federal Nuclear Center (VNIIEF), Sarov, Russia
\item \Idef{org107}Saha Institute of Nuclear Physics, Homi Bhabha National Institute, Kolkata, India
\item \Idef{org108}School of Physics and Astronomy, University of Birmingham, Birmingham, United Kingdom
\item \Idef{org109}Secci\'{o}n F\'{\i}sica, Departamento de Ciencias, Pontificia Universidad Cat\'{o}lica del Per\'{u}, Lima, Peru
\item \Idef{org110}Shanghai Institute of Applied Physics, Shanghai, China
\item \Idef{org111}St. Petersburg State University, St. Petersburg, Russia
\item \Idef{org112}Stefan Meyer Institut f\"{u}r Subatomare Physik (SMI), Vienna, Austria
\item \Idef{org113}SUBATECH, IMT Atlantique, Universit\'{e} de Nantes, CNRS-IN2P3, Nantes, France
\item \Idef{org114}Suranaree University of Technology, Nakhon Ratchasima, Thailand
\item \Idef{org115}Technical University of Ko\v{s}ice, Ko\v{s}ice, Slovakia
\item \Idef{org116}Technische Universit\"{a}t M\"{u}nchen, Excellence Cluster 'Universe', Munich, Germany
\item \Idef{org117}The Henryk Niewodniczanski Institute of Nuclear Physics, Polish Academy of Sciences, Cracow, Poland
\item \Idef{org118}The University of Texas at Austin, Austin, Texas, United States
\item \Idef{org119}Universidad Aut\'{o}noma de Sinaloa, Culiac\'{a}n, Mexico
\item \Idef{org120}Universidade de S\~{a}o Paulo (USP), S\~{a}o Paulo, Brazil
\item \Idef{org121}Universidade Estadual de Campinas (UNICAMP), Campinas, Brazil
\item \Idef{org122}Universidade Federal do ABC, Santo Andre, Brazil
\item \Idef{org123}University College of Southeast Norway, Tonsberg, Norway
\item \Idef{org124}University of Cape Town, Cape Town, South Africa
\item \Idef{org125}University of Houston, Houston, Texas, United States
\item \Idef{org126}University of Jyv\"{a}skyl\"{a}, Jyv\"{a}skyl\"{a}, Finland
\item \Idef{org127}University of Liverpool, Liverpool, United Kingdom
\item \Idef{org128}University of Tennessee, Knoxville, Tennessee, United States
\item \Idef{org129}University of the Witwatersrand, Johannesburg, South Africa
\item \Idef{org130}University of Tokyo, Tokyo, Japan
\item \Idef{org131}University of Tsukuba, Tsukuba, Japan
\item \Idef{org132}Universit\'{e} Clermont Auvergne, CNRS/IN2P3, LPC, Clermont-Ferrand, France
\item \Idef{org133}Universit\'{e} de Lyon, Universit\'{e} Lyon 1, CNRS/IN2P3, IPN-Lyon, Villeurbanne, Lyon, France
\item \Idef{org134}Universit\'{e} de Strasbourg, CNRS, IPHC UMR 7178, F-67000 Strasbourg, France, Strasbourg, France
\item \Idef{org135} Universit\'{e} Paris-Saclay Centre d¿\'Etudes de Saclay (CEA), IRFU, Department de Physique Nucl\'{e}aire (DPhN), Saclay, France
\item \Idef{org136}Universit\`{a} degli Studi di Foggia, Foggia, Italy
\item \Idef{org137}Universit\`{a} degli Studi di Pavia, Pavia, Italy
\item \Idef{org138}Universit\`{a} di Brescia, Brescia, Italy
\item \Idef{org139}Variable Energy Cyclotron Centre, Homi Bhabha National Institute, Kolkata, India
\item \Idef{org140}Warsaw University of Technology, Warsaw, Poland
\item \Idef{org141}Wayne State University, Detroit, Michigan, United States
\item \Idef{org142}Westf\"{a}lische Wilhelms-Universit\"{a}t M\"{u}nster, Institut f\"{u}r Kernphysik, M\"{u}nster, Germany
\item \Idef{org143}Wigner Research Centre for Physics, Hungarian Academy of Sciences, Budapest, Hungary
\item \Idef{org144}Yale University, New Haven, Connecticut, United States
\item \Idef{org145}Yonsei University, Seoul, Republic of Korea
\end{Authlist}
\endgroup